\documentclass[letterpaper,twocolumn,10pt]{article}
\makeatletter
\@namedef{ver@breakurl.sty}{}%
\@namedef{ver@cite.sty}{}%
\makeatother
\usepackage{usenix2019_v3}
\usepackage[sort,numbers]{natbib}
\date{}
\newcommand{\authormark}[1]{\textsuperscript{#1}}
\author{
{\rm Cesar~Pereida~Garc{\'i}a\authormark{1}, Sohaib~ul~Hassan\authormark{1}, Nicola~Tuveri\authormark{1}, Iaroslav~Gridin\authormark{1},}\\
{\rm Alejandro~Cabrera~Aldaya\authormark{1,2}, and Billy~Bob~Brumley\authormark{1}}\\
\authormark{1}Tampere University, Tampere, Finland\\
\{cesar.pereidagarcia,n.sohaibulhassan,nicola.tuveri,iaroslav.gridin,billy.brumley\}@tuni.fi\\
\authormark{2}Universidad Tecnol\'ogica de la Habana (CUJAE), Habana, Cuba\\
aldaya@gmail.com
}

\usepackage[T1]{fontenc}
\usepackage{graphicx}
\usepackage{xspace}
\usepackage{lipsum}
\usepackage{amsmath}
\usepackage{multirow}

\newcommand{\Paragraph}[1]{\smallbreak\noindent\textbf{#1.}}

\newcommand{\fl}{\textsc{Flu\-sh+\allowbreak Re\-load}\xspace}
\newcommand{\ff}{\textsc{Flu\-sh+\allowbreak Flu\-sh}\xspace}
\newcommand{\pp}{\textsc{Pri\-me+\allowbreak Pro\-be}\xspace}
\newcommand{\et}{\textsc{E\-vict+\allowbreak Ti\-me}\xspace}
\newcommand{\code}[1]{\texttt{#1}\xspace}
\newcommand{\mLibGCD}{\code{mbedtls\_mpi\_gcd}}
\newcommand{\mLibInv}{\code{mbedtls\_\-mpi\_\-inv\_\-mod}}
\newcommand{\mLibExp}{\code{mbedtls\_mpi\_exp\_mod}}
\newcommand{\pkcsOne}{PKCS~\#1\xspace}
\newcommand{\asnOne}{\code{ASN.1}\xspace}
\newcommand{\bnmodexp}{\code{BN\_mod\_exp\_mont}}
\newcommand{\bnmodinv}{\code{BN\_mod\_inverse}}
\newcommand{\flag}{\code{BN\_FLG\_CONSTTIME}}
\newcommand{\twodots}{\mathinner {\ldotp \ldotp}}
\DeclareMathOperator{\abs}{abs}

\newcommand{\footurl}[1]{\footnote{\url{#1}}}
\newcommand{\rfc}[1]{\href{https://tools.ietf.org/html/rfc#1}{RFC #1}~\citep{rfc:#1}}
\newcommand{\CVE}[1]{\href{https://cve.mitre.org/cgi-bin/cvename.cgi?name=CVE-#1}{\mbox{CVE-#1}}}
\newcommand{\cveours}{\CVE{2019-1547}\xspace}

\newcommand{\openssl}{OpenSSL\xspace}
\newcommand{\mbed}{mbedTLS\xspace}
\newcommand{\osslver}{1.1.1\xspace}
\newcommand{\opensslver}{\openssl{} \osslver{}}
\newcommand{\naf}{NAF\xspace}
\newcommand{\wnaf}{$w$\naf\xspace}
\newcommand{\nistp}{NIST~P-256\xspace}
\newcommand{\bitlen}{bitlength\xspace}

\DeclareMathOperator{\lcm}{lcm}

\title{Certified Side Channels}

\begin{document}

\maketitle

\begin{abstract}
We demonstrate that the format in which private keys are persisted impacts Side
Channel Analysis (SCA) security. Surveying several widely deployed software
libraries, we investigate the formats they support, how they parse these keys,
and what runtime decisions they make. We uncover a combination of weaknesses and
vulnerabilities, in extreme cases inducing completely disjoint multi-precision
arithmetic stacks deep within the cryptosystem level for keys that otherwise
seem logically equivalent. Exploiting these vulnerabilities, we design and
implement key recovery attacks utilizing signals ranging from electromagnetic
(EM) emanations, to granular microarchitecture cache timings, to coarse
traditional wall clock timings.
 \end{abstract}

\section{Introduction} \label{sec:intro}

Academic SCA tends to focus on implementations of cryptographic primitives in
isolation. With this view, the assumption is that any higher level protocol or
system built upon implementations of these primitives will naturally benefit
from SCA mitigations in place at lower levels.

Our work questions this assumption, and invalidates it with several concrete
vulnerabilities and attacks against modern software libraries:
we dub these \emph{Certified Side Channels}, since the novel attack vector is
deeply rooted in cryptography standards.
For this vector, ``certified'' is in the certificate sense (e.g.\ X.509), not in
the Common Criteria sense.
Counter-intuitively, we demonstrate that the format in which keys are stored
plays a significant role in real world SCA security. Detailed security
recommendations for key persistence are scarce; e.g.\ FIPS 140-2 vaguely states
``\textit{Cryptographic keys stored within a cryptographic module shall be stored either
in plaintext form or encrypted form [$\twodots$] Documentation shall specify the
key storage methods employed by a cryptographic module}'' \cite[4.7.5]{fips:140}.

There are (at least) two
high level dimensions at play regarding key formats as an SCA attack vector:
(i) Among the multitude of standardized cryptographic key formats to choose from
when persisting keys: \emph{which one to choose, and does the choice matter?}
Surprisingly, it does---we demonstrate different key formats trigger different
behavior within software libraries, permeating all the way down to the low level
arithmetic for the corresponding cryptographic primitive.
(ii) At the specification level, alongside required parameters, standardized key
formats often contain optional parameters: \emph{does including or excluding optional
parameters impact security?} Surprisingly, it does. We demonstrate that omitting
optional parameters can cause extremely different execution flows deep within a
software library, and also that two keys seemingly mathematically identical at
the specification level can be treated by a software library as inequivalent,
again reaching very different arithmetic code deep within the library.

Furthermore, we demonstrate that key parsing in general is a lucrative SCA
attack vector. This is due mostly to software engineering constraints. Complex
libraries inevitably stray to convoluted data structures containing generous
nesting levels to meet the demands of broad standardized cryptography. This is
exacerbated by the natural urge to handle keys generically when faced with
extremely diverse cryptographic standards spanning RSA, DSA, ECDSA, Ed25519, Ed448, GOST, SM2, etc.\
primitives. The motivation behind this generalization is to abstract away
underlying cryptographic details from application developers linking against a
library---more often than not, these developers are not cryptography experts.
Nevertheless, we observe that when loading keys modern security libraries make
varying design choices that ultimately impact SCA security. From the
functionality perspective, these design choices are sensible; from the security
perspective, we demonstrate they are often questionable.

\Paragraph{Outline}
\autoref{sec:background} gives an overview of the related background and previous
work. \autoref{sec:vuln} discusses the vulnerabilities discovered as a result of
our analysis, with microarchitecture SCA evaluations on \openssl RSA, DSA,
and \mbed RSA. We also demonstrate end-to-end attacks on \openssl ECDSA using timing
and EM side channels in \autoref{sec:attacks}. We conclude in \autoref{sec:conclusion}.
\section{Background} \label{sec:background}

\subsection{Public Key Cryptography}

\Paragraph{ECDSA}
Denote an order-$n$ generator $G \in E$ of an elliptic curve group $E$ with
cardinality $fn$ and $n$ a large prime and $f$ the small cofactor.
The user's private key $\alpha$ is an integer uniformly chosen from
$\{1 \twodots n-1\}$ and the corresponding public key is $D = [\alpha]G$.
With approved hash function $\textrm{Hash}()$, the ECDSA digital
signature $(r,s)$ on message $m$ (denoting with $h < n$ the
representation of $ \textrm{Hash}(m)$ as an integer) is
\begin{equation} \label{eq:ecdsa}
r = ([k]G)_x \bmod n, \quad s = k^{-1} (h + \alpha r) \bmod n
\end{equation}
where $k$ is a nonce chosen uniformly from $\{1 \twodots n-1\}$.

\Paragraph{RSA}
According to the \pkcsOne v2.2 standard (\rfc{8017}),
an RSA private key consists of the eight parameters
$\{N, e, p, q, d, d_p, d_q, i_q \}$
where all but the first two are secret, and $N=pq$ for primes $p$, $q$.
Public exponent $e$ is usually small and the following holds:
\begin{equation} \label{eq:d}
d = e^{-1} \bmod \lcm(p-1, q-1)
\end{equation}
In addition, Chinese Remainder Theorem (CRT) parameters are stored for speeding
up RSA computations:
\begin{equation} \label{eq:crt_parameters}
d_p = d \bmod p, \quad d_q = d \bmod q, \quad i_q = q^{-1} \bmod p
\end{equation}

\subsection{Key Formats}

Interoperability among different software and hardware platforms in
handling keys and other cryptographic objects requires common standards
to serialize and deserialize such objects.
\asnOne{} or \emph{Abstract Syntax Notation One} is an interface
description language to define data structures and their
(de/)serialization, standardized~\citep{temp:asnOne} jointly by ITU-T and
ISO/IEC since 1984 and widely adopted.
It supports several encoding rules, among which the \emph{Distinguished
Encoding Rules} (DER), a binary format ensuring uniqueness and
concision, has been preferred for the representation of cryptographic
objects.
\code{PEM} (\rfc{7468}) is a textual file format to store and trasmit
cryptographic objects, widespread despite being originally developed as
part of the now obsoleted IETF standards for \emph{Privacy-Enhanced
Mail} after which it is named.
\code{PEM} uses \code{base64} to encode the binary DER serialization of
an object, providing some degree of human readability and support for
text-based protocols like e-mail and HTTP(S).

\Paragraph{Object Identifiers}
The \asnOne syntax also defines an \code{OBJECT~IDENTIFIER} primitive
type which represents a globally unique identifier for an object. ITU-T
and ISO jointly manage a decentralized hierarchical registry of object
identifiers or \code{OID}s. The registry is organized as a tree
structure, where every node is authoritative for its descendants, and
decentralization is obtained delegating the authority on subtrees to
entities such as countries and organizations. This mechanism solves the
problem of assigning globally unique identifiers to entities to
facilitate global communication.

\Paragraph{RSA private keys}
\pkcsOne (\rfc{8017}) also defines the \asnOne DER encoding for an RSA
private key, defining an item for each of its eight parameters.
As further discussed in \autoref{sec:mbed:RSA:missing_params},
the standard does not strictly require implementations to include all
the eight parameters during serialization, nor to invalidate the object
during deserialization if one of the parameters is not included.

\Paragraph{EC private keys}
The ANSI X9.62 standard \citep{temp:ansi:X9.62:2005} is the normative
reference for the definition of the ECDSA cryptosystem and the encoding
of ECDSA public keys, but omits a serialization for private keys.
The SEC1 standard~\citep{sec1} follows ANSI X9.62 for the public key
\asnOne and provides a DER encoding also for EC private keys, but allows
generous variation as it seems to assume different encapsulating options
depending on different protocols in which the EC private key can be
used.
Flexibility in the format brings complexity in the deserializer
implementation, that needs to be stateful w.r.t. parsing of the
container of the private key encoding and flexible enough to
interoperate with other implementations and interpretations of the
standards: this already suggests that the parsing stage shows potential
as a lucrative SCA attack vector.
The SEC1 \asnOne notation for \code{ECPrivateKey} contains the private
scalar as an octet string, an optional (depending on the container)
\code{ECDomainParameters} field, and an optional bit string field to
include the public part of the key pair.
The \code{ECDomainParameters} can be null, if the curve
parameters are specified in the container encapsulating the
\code{ECPrivateKey}, or contain either an \code{OID} for a
``named'' curve, or a \code{SpecifiedECDomain} structure.
The latter, simplifying, contains a description of the field over which
the EC group is defined, the definition of the curve equation in terms
of the coefficients of its Short Weierstrass form, an encoding of the
EC base point, and its order $n$.
Finally it can \emph{optionally} contain a component to represent a
small cofactor $f$ as defined at the beginning of this section.
In \autoref{sec:expl_params_vuln} we will further discuss about the security
consequences caused in actual implementations by the logic required to
support the cofactor as an optional field.

\Paragraph{MSBLOB key format}
MSBLOB is the \openssl{} implementation of Microsoft's private key BLOB format%
\footurl{https://docs.microsoft.com/en-us/windows/win32/seccrypto/base-provider-key-blobs}
supporting different cryptosystems, using custom defined structures and fields.
DSS key BLOB uses an arbitrary structure, while RSA key BLOBs follows \pkcsOne
with minor differences.
To identify each cryptosystem, a ``magic member'' is used in the key BLOB
structure---the member is the hexadecimal representation of the ASCII encoding of the
cryptosystem name, e.g.\ ``RSA1'', ``RSA2'', ``DSS1'', ``DSS2'', etc., where
the integer dictates if it is a public or a private key.
Public and private key BLOBs are stored as binary files in little-endian order and
by default the private key BLOBs are not encrypted---it is up to the developers
to choose whether to encrypt the key. Microsoft created the public and private
key BLOBs in order to support cryptographic service providers (CSP),
i.e.\ third party cryptographic software modules.
It is worth noting that both private and public BLOBs are independent from each
other, thus allowing a CSP to only support and implement the desired format
according to the cryptosystem in use, meaning that public keys can be computed
on-demand using the private key BLOB information.

\Paragraph{PVK key format}
The PriVate Key (PVK) format is a Microsoft proprietary
key format used in Windows supporting signature generation using both DSA and RSA
private keys. Little information is available about this format but a key is typically
composed of a header containing metadata, and a body containing a private key BLOB
structure as per the previous description.
Following the same idea as in the private key BLOB, the PVK header metadata contains
the ``magic'' value \code{0xb0b5f11e}\footnote{Leetspeak for ``bobsfile''!} to
uniquely identify this key format.
Additionally, PVK's header contains metadata information for key password protection,
preventing the storage of private key information in plain text. Unfortunately, PVK
is an outdated format and it only supports RC4 encryption, moreover, in some cases
PVK keys use a weakened encryption key to comply with the US export restrictions
imposed during the 90's \footurl{http://justsolve.archiveteam.org/wiki/PVK}.

\subsection{Side-Channel Analysis} \label{sec:sca}

SCA is a cryptanalysis technique used to target software
and hardware implementations of cryptographic primitives. The main goal of SCA is
to expose hidden algorithm state by measuring variations in time, power consumption,
electromagnetic radiation, temperature, and sound. These variations might leak
data or metadata that allows the retrieval of confidential information such as
private keys and passwords.
The history of SCA is long and rich---from the military program called TEMPEST \citep{1972:tempest}
to current commodity PCs, SCA has deeply impacted security-critical systems and
it has reached the most popular and widely used cryptosystems over the years such
as AES, DSA, RSA, and ECC, implemented in the most widely used cryptographic
libraries including \openssl, BoringSSL, LibreSSL, and \mbed.

SCA can be broadly categorized (w.r.t.\ signal procurement techniques) in two specific research fields: hardware and software.
Both fields have evolved and developed their own techniques, and the line separating
them has blurred as research improves, and attacks become more complex.
Nevertheless, the ultimate goal is still the same: extract confidential information
from a device executing vulnerable cryptographic code. A brief overview follows.

\Paragraph{Hardware}
Ever since their inception, System-on-Chip (SoC) embedded devices
have become passively ubiquitous in the form of mobile devices and IoT,
performing security critical tasks over the Internet. Their basic building
blocks---in terms of performing computations---are the CMOS transistors,
drawing current during the switching activity to depict the behavior of logic
gates. Power analysis attacks introduced by \citet{DBLP:conf/crypto/KocherJJ99} rely on the
fact that accumulated switching activity of these transistors
influence the overall power fluctuations while secret data dependent computations
take place on the processor and memory subsystems.

While power analysis is one way to perform SCA, devices
may also leak sensitive information through other means such as
EM \citep{DBLP:conf/ches/AgrawalARR02}, acoustic \citep{DBLP:conf/crypto/GenkinST14},
and electric potential \citep{DBLP:journals/jce/GenkinPT15}.
In contrast to the power side channels which require physically tapping onto the power lines,
EM and acoustic based SCA add a spatial dimension. There may be
slight differences when it comes to acquiring and processing these signals, but
in essence the concept is similar to traditional power analysis, hence the
hardware based SCA techniques generally apply to all.

Over the years more powerful SCA techniques have emerged
such as differential power analysis \citep{DBLP:conf/crypto/KocherJJ99},
correlation power analysis \citep{DBLP:conf/ches/BrierCO04},
template attacks \citep{DBLP:conf/ches/ChariRR02}, and horizontal attacks \citep{BauerJPRW15}.
Most of these techniques rely on statistical methods to find small secret data dependent leakages.

Traditionally, hardware SCA research mainly focuses on architecturally simpler devices
such as smart cards and microcontrollers
\citep{DBLP:conf/esmart/QuisquaterS01,DBLP:journals/tc/MessergesDS02,DBLP:journals/dt/PoppMO07}.
Being simple here does not imply that developing and deploying such cryptosystems is simpler, rather
in terms of their functionality and hardware architecture.
Modern consumer electronics (e.g. smart phones) are more feature rich, containing SoC
components, memory subsystems and multi-core processors with clock speeds in gigahertz. These devices
are often running a full operating system (several in fact)
making it possible to deploy software libraries such as \openssl. More recently,
a new class of hardware side channel attacks on embedded, mobile devices and even PCs has emerged,
targeting crypto software libraries such as
\openssl \citep{DBLP:conf/ccs/GenkinPPTY16,DBLP:conf/ches/GaleaMPT15},
GnuPG \citep{DBLP:conf/crypto/GenkinST14,DBLP:conf/ctrsa/GenkinPPT16,DBLP:journals/jce/GenkinPT15,DBLP:conf/ches/GenkinPPT15},
PolarSSL \citep{DBLP:conf/cosade/DugardinPNBDG16},
Android's Bouncy Castle \citep{DBLP:conf/ctrsa/BelgarricFMT16},
and WolfSSL \cite{DBLP:conf/ctrsa/SamwelBBDS18}.
They employ various signal processing tools to counter the noise induced by
complex systems and microarchitectures.
For further details, \citet{DBLP:books/sp/17/Tunstall17} present an elaborate
discussion on hardware based SCA techniques,
while \citet{DBLP:journals/jce/DangerGHMN13} and \citet{DBLP:journals/iacr/AbarzuaCL19} sum
up various SCA attacks and their countermeasures.

\Paragraph{Software}
The widespread use of e-commerce and the need for security on the Internet
sparked the development of cryptographic libraries such as \openssl.
Researchers quickly began analyzing these libraries and it took a short time
to find security flaws in these libraries. Impulsed by Kocher's work \citep{DBLP:conf/crypto/Kocher96},
SCA timing attacks quickly gained traction. By measuring the amount of time
required to perform private key operations, the author demonstrated that
it was feasible to find Diffie-Hellman exponents, factor RSA keys, and recover DSA keys.
Later \citet{DBLP:conf/uss/BrumleyB03} demonstrated that it was possible to do
the same but remotely, by measuring the response time from an \openssl-powered web
server.
Other TLS-level timing attacks include \citep{DBLP:conf/ches/KlimaPR03} with a
software target and \citep{DBLP:conf/uss/MeyerSWSST14} with a hardware target.

As software SCA became more complex and sophisticated, a new subclass of
attacks denominated ``microarchitecture attacks'' emerged.
Typically, a modern CPU executes multiple programs either concurrently or via
time-sharing, increasing the need to optimize resource utilization to obtain
high performance. To achieve this goal, microarchitecture components try to
predict future behavior and future resource usage based on past program states.
Based on these observations, researchers \citep{2005:bernstein,Percival05} discovered
that some microarchitecture components---such as the memory subsystem---work wonderfully
as communication channels. Due to their shared nature between programs, some of
the microarchitecture components can be used to violate access control and achieve
inter-process communication.
Among these components, researchers noticed that the memory subsystem is arguably
the easiest to exploit: by observing the memory footprint an attacker can leak
algorithm state from an executing cryptographic library in order to obtain secret keys.
Since the initial discovery, several SCA techniques have been developed
to extract confidential data from different memory levels and under different
threat models.
Some of these techniques include \fl \citep{DBLP:conf/uss/YaromF14},
\pp \citep{DBLP:conf/ctrsa/OsvikST06}, \et \citep{DBLP:conf/ctrsa/OsvikST06},
and \ff \citep{DBLP:conf/dimva/GrussMWM16}.
Moreover, recent research \citep{temp_portsmash,temp_ridl,temp_fallout} shows that
most (if not all) microarchitecture components shared among programs are a security
hazard since they can potentially be used as side-channels.
\citet{DBLP:journals/jce/GeYCH18} provide a great overview on software SCA,
including the types of channels, microarchitecture components, side-channel attacks,
and mitigations.

\subsection{Lattice Attacks}\label{sec:bg:lattice}

In \autoref{sec:attacks} we present two attacks against ECDSA signing
that differ in SCA technique, but share a common pattern:
(i) gathering several $(r, s, m)$ tuples in a collection phase,
using SCA to infer partial knowledge about the nonce used
during signature generation;
(ii) a recovery phase combines the collected tuples and the associated
partial knowledge to retrieve the long-term secret key.

To achieve the latter, we recur to the common strategy of constructing
\emph{hidden number problem} (HNP)~\citep{DBLP:conf/crypto/BonehV96}
instances from the collected information, and then use lattice
techniques to find the secret key.
In this section we discuss the lattice technique used to recover the
private keys.

We follow the formalization used in \citep{DBLP:conf/uss/GarciaB17},
which itself builds on the work by
\citet{DBLP:journals/joc/NguyenS02,DBLP:journals/dcc/NguyenS03}, that
assumed a fixed amount of known bits (denoted $\ell$) for each nonce
used in the lattice, but also includes the improvements by
\citet{DBLP:conf/ches/BengerPSY14}, using $\ell_i$ and $a_i$ to
represent, respectively, the amount of known bits and their value on a
per-equation basis. 

The collection phase of \citep{DBLP:conf/uss/GarciaB17} as well as our
\autoref{sec:attacks:EM} attack recovers
information regarding the LSBs of each
nonce, hence it annotates the nonce associated with $i$-th equation as
$k_i = W_i b_i + a_i$,
with $W_i = 2^{\ell_i}$,
where $\ell_i$ and $a_i$ are known, and since
$0 < k_i < n$
it follows that
$0 \leq b_i \leq n/W_i$.
Denote $\lfloor x \rfloor_n$ modular reduction of $x$ to the interval $\{0 \twodots n-1\}$
and $\lvert x \rvert_n$ to the interval $\{-(n-1)/2 \twodots (n-1)/2\}$.
Combining \eqref{eq:ecdsa}, define (attacker-known) values
$t_i = \lfloor r_i/(W_i s_i) \rfloor_n$ and $\hat{u}_i = \lfloor (a_i-h_i/s_i)/W_i \rfloor_n$,
then $0 \leq \lfloor \alpha t_i - \hat{u}_i \rfloor_n < n/W_i$ holds.
Setting $u_i = \hat{u}_i + n/2W_i$ we obtain $v_i = \lvert \alpha t_i - u_i \rvert_n \leq n/2W_i$,
i.e.\ integers $\lambda_i$ exist such that
$\abs(\alpha t_i - u_i - \lambda_i n) \leq n/2W_i$
holds.
Thus $u_i$ approximate $\alpha t_i$ since they are closer than a
uniformly random value from $\{1 \twodots n-1\}$, leading to an instance of the
HNP~\citep{DBLP:conf/crypto/BonehV96}: recover $\alpha$ given many $(t_i,u_i)$
pairs.

Consider the rational $d+1$-dimension lattice generated by
the rows of the following matrix.
\[
B =
\begin{bmatrix}
2 W_1 n & 0 & \dots & \dots & 0 \\
0 & 2 W_2 n & \ddots & \vdots & \vdots \\
\vdots & \ddots & \ddots & 0 & \vdots \\
0 & \dots & 0 & 2 W_d n & 0 \\
2 W_1 t_1 & \dots & \dots & 2 W_d t_d & 1
\end{bmatrix}
\]
Denoting $\vec{x}=(\lambda_1,\ldots,\lambda_d,\alpha)$,
$\vec{y}=(2 W_1 v_1,\ldots,2 W_d v_d,\alpha)$, and
$\vec{u}=(2 W_1 u_1,\ldots,2 W_d u_d,0)$,
then $\vec{x}B-\vec{u}=\vec{y}$ holds.
Solving the Closest Vector Problem (CVP) with inputs $B$ and $\vec{u}$ yields
$\vec{x}$, and hence the private key $\alpha$.
Finally, as in \citep{DBLP:conf/uss/GarciaB17},
we embed the CVP into a Shortest Vector Problem (SVP) using the
classical strategy \cite[Sec.\ 3.4]{DBLP:conf/crypto/GoldreichGH97a},
and employ an extended search space heuristic
\cite[Sec.\ 5]{DBLP:conf/eurocrypt/GamaNR10}.

The presence of outliers among the results of the collection phase
usually has a detrimental effect on the chances of success of the
lattice attack. The traditional solution is to oversample, filtering
$t > d$ traces from the collection phase if $d$ traces are required to
embed enough leaked information in the lattice instance to solve the HNP.
Indicating with $e$ the amount of traces with errors in the filtered set
of size $t$, picking a subset of size $d$ uniformly at random, the
probability for any such subset to be error-free is
$\hat{p} =\left. \binom{t-e}{d} \middle/ \binom{t}{d} \right.$.
For typical values of $\{t, e, d\}$, $\hat{p}$ will be small. Viewing the
process of randomly picking a subset and attempting to solve the resulting
lattice instance as a Bernoulli trial, the number of expected trials before
first success is $1/\hat{p}$. So an attacker can compensate for small $\hat{p}$
by running $j=1/\hat{p}$ jobs in parallel.

\subsection{Triggerflow} \label{sec:tf}
Triggerflow \citep{DBLP:conf/dimva/GridinGTB19} is a tool for tracking execution paths,
previously used to facilitate SCA of \openssl. After users mark up source code
with annotations of Points Of Interest (POI) and filtering rules for false positive considerations, Triggerflow runs
the binary executable under a debugger and records the execution paths that led up to POIs.
The user supplies binary invocation lines called ``triggers''.
These techniques are useful in SCA of software, where areas that do
not execute in constant time are known and the user needs to find code that leads up
to them. The authors designed Triggerflow with continuous integration (CI) in mind, and maintain an
automatic testing setup which continuously monitors all non-EOL branches of \openssl for new vulnerabilities by
watching execution flows that enter known problematic areas.

Triggerflow is intended for automated regression testing and has no
support for automatic POI detection. Thus offensive leakage detection
methodologies including (but certainly not limited to)
CacheAudit \cite{DBLP:journals/tissec/DoychevKMR15},
templating \cite{DBLP:conf/asiacrypt/BrumleyH09,DBLP:conf/uss/GrussSM15},
CacheD \cite{DBLP:conf/uss/WangWLZW17},
and DATA \cite{DBLP:conf/uss/WeiserZSMMS18}
complement Triggerflow to establish POIs. One approach is to apply these leakage
detection methodologies, filter out false positives, limit to functions deemed
security-critical and worth tracking, then use the result to add Triggerflow
source code annotations for CI. See \cite[Sec.~7]{DBLP:conf/dimva/GridinGTB19}
for a more extensive discussion.
\section{Vulnerabilities} \label{sec:vuln}

\begin{figure}
\includegraphics[width=\linewidth]{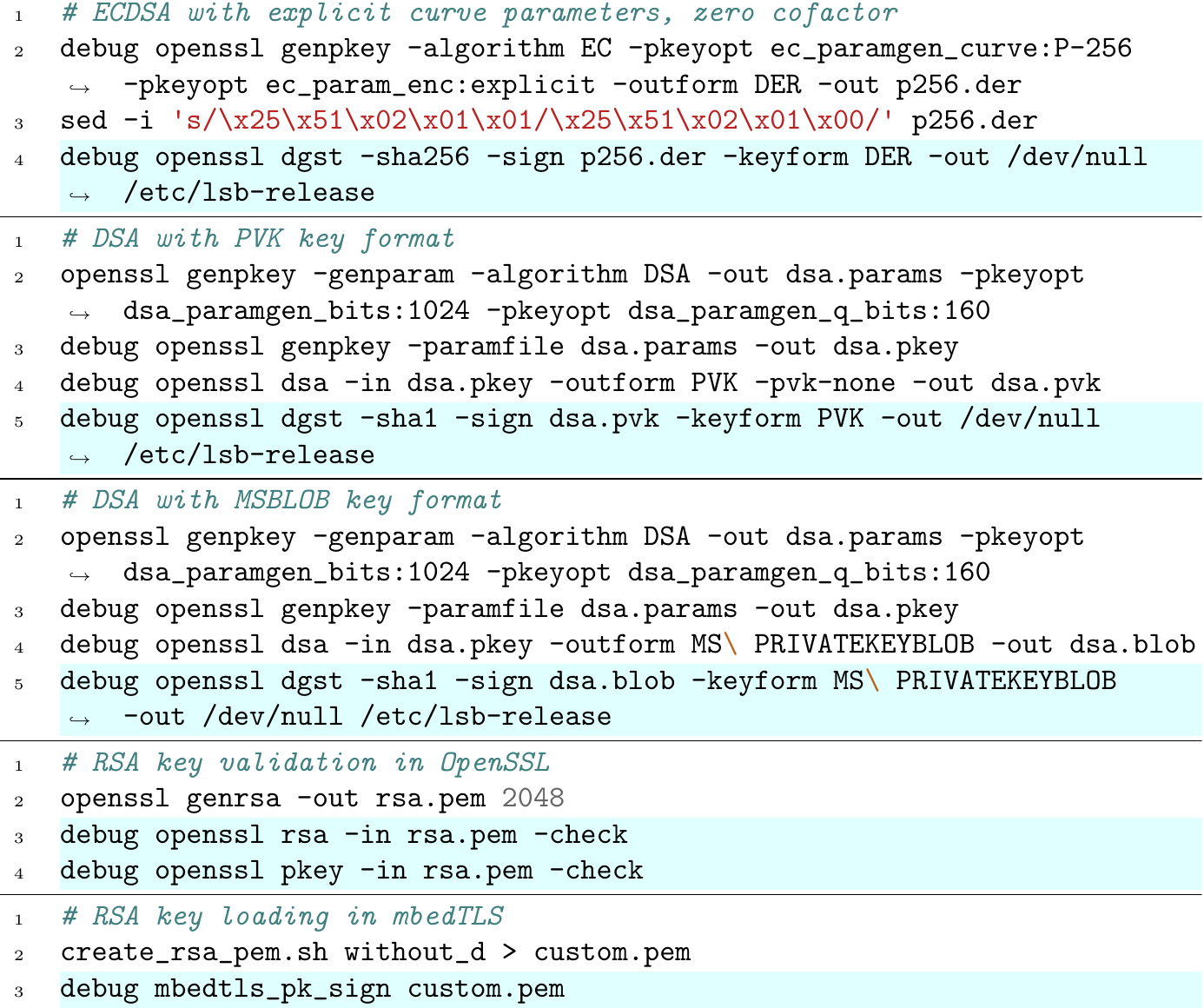}
\caption{New Triggerflow unit tests.}
\label{fig:triggers}
\end{figure}

We used Triggerflow to analyze several code paths on multiple cryptographic
libraries, discovering SCA vulnerabilities across \openssl and \mbed. In this
section, we discuss these vulnerabilities, including the unit tests we developed
for Triggerflow that detected each of them, then
identify the root cause in each case.
Following \autoref{fig:triggers}, Triggerflow executes each line of the unit
tests given in a text file. Triggerflow will trace the execution of lines
beginning with \code{debug} to detect break points getting hit at SCA-critical
points in the code. Each such line is security critical---in these examples,
generating a key pair or using the private key to e.g.\ digitally sign a message. Hence
if Triggerflow encounters said break points during execution, it represents a
potential SCA vulnerability.
We compiled the target executables (and shared libraries) with debug symbols,
and source code annotated using Triggerflow's syntax to mark previously known
SCA-vulnerable functions. Lines that do not begin with \code{debug} are not
traced by Triggerflow, merely executed as preparation steps for subsequent
triggers (e.g.\ setting up public fixed parameters).

Vulnerability-wise, the main results of this section are as follows:
(i) bypassing SCA countermeasures using ECC explicit parameters (\autoref{sec:expl_params_vuln}, \openssl);
(ii) bypassing SCA countermeasures for DSA using PVK and MSBLOB key formats (\autoref{sec:dsa_openssl}, \openssl);
(iii) bypassing SCA countermeasures for RSA by invoking key validation (\autoref{sec:rsa_openssl}, \openssl);
(iv) bypassing SCA countermeasures for RSA through key loading (\autoref{sec:mbed:RSA:missing_params}, \mbed).

\subsection{ECC: Bypass via Explicit Parameters}\label{sec:expl_params_vuln}

From a standardization perspective, curve data for ECC key material gets
persisted in one of two ways: either including the specific OID that points to a
named curve with fixed parameters, or explicitly specifying the curve
with \asnOne{}
syntax. Mathematically, they seem equivalent. To explore the potential
difference in security implications between these options, we constructed three
keys:
(i) a \nistp private key as a named curve, using the
\code{ec\_param\_enc:named\_curve} argument to the \openssl \code{genpkey}
utility;
(ii) a \nistp private key with explicit curve parameters, using the
\code{ec\_param\_enc:explicit} argument;
(iii) a copy of the previous key, but post-modified with the \openssl
\code{asn1parse} utility to remove the optional cofactor.
The first two keys additionally used the \code{ec\_paramgen\_curve:P-256}
argument to specify the target curve. We highlight that, from a standards
perspective, all three of these keys are valid. We then integrated the commands
to produce these keys into the Triggerflow framework as unit tests. Finally, we
added an \openssl \code{dgst} utility unit test for each of these keys in
Triggerflow, to induce ECDSA signing. What follows is a discussion on the three
distinct control flow cases for each key, regarding the security-critical scalar
multiplication operation.

\Paragraph{Named curve}
Triggerflow indicated \code{ecp\_\-nistz256\_\-points\_\-mul} handled the
operation. The reason for this is \openssl uses an \code{EC\_METHOD} structure
for legacy ECC; the assignment of structure instances to specific curves happens
at library compile time, allowing different curves to have different (optimized)
implementations depending on architecture and compiler features. This particular
function is part of the \code{EC\_GFp\_nistz256\_method}, an \code{EC\_METHOD}
optimized for AVX2 architectures \citep{DBLP:journals/jce/GueronK15}. The
implementation is constant time, hence this is the best case scenario.

\Paragraph{Explicit parameters}
Triggerflow indicated \code{ec\_\-scalar\_\-mul\_\-ladder} handled the operation,
through the default \code{EC\_GFp\_simple\_method}, the generic implementation
for curves over prime fields. In fact this is the oldest \code{EC\_METHOD} in
the codebase, present since ECC support appeared in 2001. The implementation of
this particular function was mainlined in 2018 \citep{DBLP:conf/acsac/TuveriHGB18}
as a result of \CVE{2018-5407} \citep{temp_portsmash}, SCA-hardening generic
curves with the standard Montgomery ladder. Interpreting this Triggerflow
result, we conclude \openssl has no runtime mechanism to match explicit
parameters to named curves present in the library. Ideally, it would match the
explicit parameters to \code{EC\_GFp\_nistz256\_method} for improved performance
and SCA resistance. Failure to do so bypasses one layer of SCA mitigations, but
in this particular case the default method still features sufficient SCA
hardening.

\Paragraph{Explicit parameters, no cofactor}
Triggerflow indicated \code{ec\_\-wNAF\_\-mul} handled the operation through the
same \code{EC\_METHOD} as the previous case. This is a known SCA-vulnerable
function since 2009 \citep{DBLP:conf/asiacrypt/BrumleyH09}, and is a POI
maintained in the Triggerflow patchset to annotate \openssl for automated CI.
Root causing the failed Triggerflow unit test, the function only early exits to
the SCA-hardened ladder if both the curve generator order and the curve
cardinality cofactor are non-zero. Since the optional cofactor is not present in
the key, the library assigns zero as the default, indicating either the provided
cofactor was zero or not provided at all. The \openssl ladder implementation
utilizes the cofactor as part of SCA hardening, hence the code unfortunately
falls through to the SCA-insecure version in this case, bypassing the last layer
of SCA defenses for scalar multiplication. This is the path we will exploit in
\autoref{sec:attacks}.

\Paragraph{Keys in the wild}
While we reached a vulnerable code path through a standards-compliant, valid,
non-malicious key, the fact is the \openssl CLI will not organically emit a key
in this form. One can argue that \openssl is far from the only security tool
that produces keys conforming to the specification, that it must subsequently
parse since they are valid. Nevertheless, this leaves us with the question:
\emph{do keys like this exist---does this vulnerability matter?} Investigating,
we at least found two deployment classes this vulnerability affects:
(i) The GOST engine\footurl{https://github.com/gost-engine/engine} for
\openssl, dynamically adding support for Russian cryptographic primitives in
\rfc{4357}. Since the curves from the standard are not built-in to \openssl, the
engine programatically constructs the curve based on fixed parameters inside the
engine. However, since the cofactor parameter to the \openssl
\code{EC\_GROUP\_set\_generator} API is optional, the engine developers omit it
in earlier versions, passing \code{NULL}. When GOST keys are persisted, they
have their own OID distinct from legacy ECC standards and only support named
curves; however, the usage of these curves within the engine hits the same exact
code path.
(ii) GOSTCoin\footurl{https://github.com/GOSTSec/gostcoin} is the official
software stack for a cryptocurrency. It links against \openssl for cryptographic
functionality, but does not support the GOST engine. Examining the digital
wallet, we manually extracted several DER-encoded legacy (OID-wise) ECC private
keys from the binary. Parsing these keys revealed they are private keys with
explicit parameters from \rfc{4357}, ``Parameter Set A''. Upon closer inspection,
the cofactor is present in the \asnOne{} encoding, yet explicitly set to zero.
Similar to the previous case, this is due to failure to supply the correct
cofactor to the \openssl EC API when constructing the curve.

From this brief study, we can conclude that failure to provide the valid
cofactor to the \openssl EC API when constructing curves programmatically (the
only choice for curves not built-in to the library), or importing a (persisted)
ECC private key with explicit parameters containing a zero or omitted
(spec-optional) cofactor are characteristics of applications affected by this
vulnerability.

\Paragraph{Related work}
Concurrent to our work, \citet{temp:2019:fault} utilize explicit parameters in
\openssl to mount a fault injection attack. They invasively induce a fault during
key parsing to change \openssl{}'s representation of a curve coefficient. This
causes decompression of the explicit generator point to emit a point on a weaker
curve, subsequently mounting a degenerate curve attack \citep{DBLP:journals/iet-ifs/NevesT18}.
At a high level, the biggest differences from our work are the invasive attack
model and limited set of applicable curves.

Subsequent to our work, \CVE{2020-0601} tracks the ``Curveball''
vulnerability. It affects the Windows CryptoAPI and uses ECC explicit parameters
to match a named curve in all but the custom generator point, allowing to spoof
code-signing certificates.

\subsection{DSA: Bypass via Key Formatting} \label{sec:dsa_openssl}

As the Swiss knife of cryptography, \openssl provides support for PVK and MSBLOB
key formats to perform digital signatures using DSA. In fact, \openssl has supported
these formats since version 1.0.0, hence the library has a dedicated
file in \texttt{crypto/pem/pvkfmt.c} for parsing these keys.
The file contains all the logic to parse Microsoft's DSA and RSA private key
BLOBs, common to both PVK and MSBLOB key formats. Unfortunately, the bulk of code
for parsing the keys has seen few changes throughout the years, and more
importantly it has missed important SCA countermeasures that other parts
of the code base have received \citep{DBLP:conf/ccs/GarciaBY16},
allowing this vulnerability to go unnoticed in all \openssl branches
until now.

As mentioned previously,
PVK and MSBLOB key files contain only private key material but \openssl expects
the public key to be readily available. Thus every time it loads any of these key formats,
the library computes the corresponding public key. More specifically,
the upper level function \texttt{b2i\_dss} reads the private key material and
subsequently calls the \texttt{BN\_mod\_exp} function to compute the public key
using the default modular exponentiation function, without first setting the constant-time
flag \texttt{BN\_FLG\_CONSTTIME}. Note that this vulnerability does not depend
on whether the PVK key is encrypted or not, because when the code reaches the
\texttt{b2i\_dss} function, the key has been already decrypted, and the modular
exponentiation function is already leaking private key material.
This default SCA-vulnerable modular exponentiation algorithm follows a
square-and-multiply approach---first pre-computes a table of multipliers,
and then accesses the table during the square-and-multiply step.
Already in 2005 \citet{Percival05} demonstrated an L1 data cache-timing attack
against this function during RSA decryption. We found that the original flaw is still present,
but this time in the context of DSA.

\begin{figure}
\includegraphics[width=\linewidth]{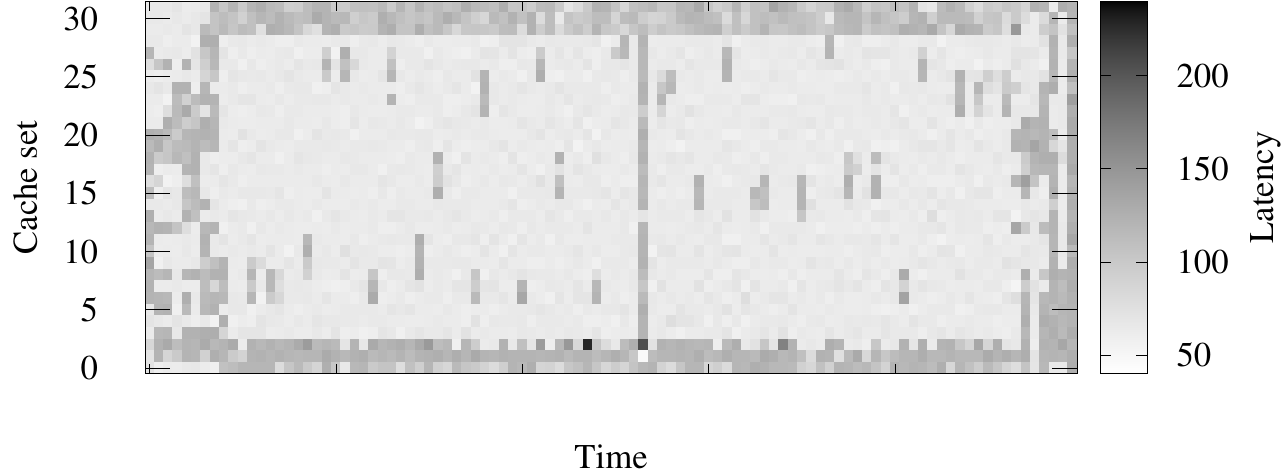}
\caption{L1 dcache trace showing distinctive access patterns to pre-computed
multipliers in cache sets 6-8, 9-11, 13-16, 15-17, 22-24, 25-27, 28-30 during
DSA public key computation.}
\label{fig:dsa}
\end{figure}

\autoref{fig:dsa} demonstrates the side-channel leakage obtained by our L1 data-cache
malicious spy process running in parallel with \openssl during a modular exponentiation
operation while computing the DSA public key using PVK and MSBLOB key formats.
Using the \pp technique, our spy process is able to measure the latency of accessing
a specific cache set (y-axis) over time (x-axis) to obtain a sequence of pre-computed
multipliers accessed during computation. In OpenSSL a multiplier is represented
as a BIGNUM structure spanning approximately across three different cache sets.
Reading from top-to-bottom and left-to-right, and after a brief period of noise,
the figure shows that every block of approximately three continuous high latency
cache sets corresponds to a multiplier access.
An attacker can not only trace the multipliers accessed, but also the order in
which they were accessed during the exponentiation, leaking more than half of
the exponent bits. This information greatly reduces the effort to perform full
key recovery. Moreover, the public key is computed every time the private key is
loaded, thus an attacker has several attempts at tracing the sequence of operations
performed during the exponentiation. Our experiments reveal that cache sets
stay constant across multiple invocations of modular exponentiation, reducing the
attacker's effort and permitting the use of statistical techniques to improve
the leakage quality.

\Paragraph{Keys in the wild}
PVK and MSBLOB are based on MS proprietary private key formats---nevertheless
they are widely found in use in open source software.
MSBLOB keys are supported by MS Smart Card CSP and
OpenSC\footnote{\url{https://github.com/OpenSC/OpenSC}},
an open source software library for smart cards linking to OpenSSL.
In fact, OpenSC has a
function\footnote{\url{https://github.com/OpenSC/OpenSC/blob/master/src/minidriver/minidriver.c\#L3308}}
that creates a key container---by calling the OpenSSL vulnerable function---whenever
``the card either does not support internal key generation or the caller requests
that the key be archived in the card'', facilitating the attack in a smart card setting.
On the other hand, MS Visual Studio 2019 provides tools%
\footnote{\url{https://docs.microsoft.com/en-us/windows-hardware/drivers/devtest/tools-for-signing-drivers}}
to generate, convert, and sign Windows drivers, libraries, and catalog files
using the PVK format.
In a typical workflow, \texttt{MakeCert} generates certificates and the
corresponding private key, then \texttt{Pvk2Pfx} encapsulates private keys and
certificates in a PKCS \#12 container, and finally \texttt{SignTool} signs the
driver.
Interestingly, \texttt{MakeCert} and \texttt{SignTool} successfully generate keys
and signatures using RSA and DSA, but \texttt{Pvk2Pfx} fails to accept any key
that is not RSA---a gap filled by the vulnerable OpenSSL, creating compliant
PKCS \#12 keys.
Other libraries such as \texttt{jsign}\footnote{\url{https://ebourg.github.io/jsign/}},
\texttt{osslsigncode}\footnote{\url{https://sourceforge.net/projects/osslsigncode/files/osslsigncode/}},
and the Mono Project\footnote{\url{https://www.mono-project.com/}} exist to provide
signing capabilities using MS proprietary private key formats outside of Windows.
We can expect this vulnerability to be exploitable by an attacker targeting Windows
developers.

\subsection{RSA: Bypass via Key Validation} \label{sec:rsa_openssl}

RSA key validation is a common operation required in a cryptography library supporting
RSA to verify that an input key is indeed a valid RSA key.
We found that \openssl function \code{RSA\_check\_key\_ex} located at
\code{crypto/rsa/rsa\_chk.c} contains several SCA vulnerabilities.
In fact, we found that the affected function \code{RSA\_check\_key\_ex}
can be accessed by two public entry points: a direct call to
\code{RSA\_check\_key}, and through the public EVP interface calling
\code{EVP\_PKEY\_check} on an RSA key.
\autoref{fig:triggers} shows the commands in \openssl leading to the affected
code path through the two different public functions. Note that any external,
\openssl{}-linking application calling any of these two public functions is also affected.

The check function takes as input an RSA key, parses the key, and reads all of the
private and public components, checking the correctness of all the components.
In general, the function validates the primality of $p$ and $q$, then it recomputes
the rest of the values $\{N, d, d_p, d_q, i_q\}$ to compare against the parsed
values and check their validity. Unfortunately, we found that in several cases \openssl
uses by default SCA-vulnerable functions to recompute these secret values.

\Paragraph{Primality testing vulnerabilities}
The prime values $p$ and $q$ are the first components
verified during the process. The verification is done using the Miller-Rabin
primality test \citep{MR566880} as implemented in the function
\code{BN\_is\_prime\_fasttest\_ex}. This function calls a lower level
\textit{witness} function named \code{bn\_miller\_rabin\_is\_prime}\footnote{In
\openssl 1.0.2 the function is called \code{witness}.}
where a $b$ base value is chosen randomly to compute $b^m \bmod p$, in
which $p$ is the candidate prime and the relation $2^{a}m = p - 1$ holds.
The \textit{witness} exponentiation is performed using the \bnmodexp function,
where unfortunately the \flag is not set beforehand. Thus a variable-time sliding
window exponentiation is used, allowing a malicious process to potentially
perform a data cache-timing attack to recover half of the bits from the
exponent \citep{Percival05}. This is enough information to recover both prime values
$p$ and $q$.
Moreover, the exponentiation function gets called several time by the \textit{witness}
function with different $b$ values in order to obtain confidence about the
prime values, providing multiple attempts for an attacker to capture the leakage
and perform error correction during its key recovery attack.

In addition to the previous vulnerability, as part of the \textit{witness}
function, a Montgomery setup phase occurs in \code{BN\_MONT\_CTX\_set}, where
the inverse of $2^w \bmod p$ for \textit{w}-bit architectures is computed.
The modular inverse function \bnmodinv is called without setting the constant-time
flag. The inverse operation uses a variation of the greatest common divisor (GCD)
algorithm, which is dependent on its inputs $\{2^w, p \bmod 2^w\}$,
thus leaking algorithm state equivalent to the least significant word of both
$p$ and $q$.

\Paragraph{Secret value vulnerabilities}
Once the prime values $p$ and $q$ are deemed correct, the key
validation continues by computing the rest of the secret components where more
vulnerabilities are found.
To compute the private exponent $d$ during the verification code path,
\openssl uses the least common multiple (LCM) of $p-1$ and $q-1$. Nevertheless,
this operation is computed as
\begin{equation} \label{eq:lcm}
    \lcm(p-1, q-1) = \frac{(p - 1) \cdot (q - 1)}{\gcd(p - 1, q - 1)}
\end{equation}
performing the GCD computation using the \code{BN\_gcd} function.
This function does not have an early exit to a constant-time function, instead
it completely ignores the flag existence, so even if it was set it would not have
any effect on the code path taken.
Finally, the last vulnerability is observed during CRT $i_{q}$ computation.
\openssl computes this parameter using the \bnmodinv function, which yet again
fails to properly set the constant-time flag, leaving the computation
$q^{-1} \bmod p$ unprotected.

It is worth noting that variable-time GCD functions, and variants, potentially leak
all the algorithm state. Depending on the attacker capabilities \citep{cryptoeprint:2020:055},
an attacker is fully capable of recovering the input values, i.e.\ $p$ and $q$.

As can be observed, all of the vulnerabilities leak on $p$ and $q$
at different degrees, but by combining all the leaks, an attacker can use the
redundancy and number-theoretic constraints to correct errors and obtain
certainty on the bits leaked.

\Paragraph{Keys in the wild}
Surprisingly, the vulnerabilities presented in this section do not depend on
a special key format. In fact, the vulnerabilities are triggered whenever an RSA
key is checked for validity using the OpenSSL library, thus a potential attacker
could simply wait for the right moment to exploit these vulnerabilities.
The potential impact of these vulnerabilities is large, but it is minimized by
two important factors: the user must trigger an RSA key validation; and the
attacker must be collocated in the same CPU as the user.
Nevertheless, this is not a rare scenario, and thus exploitation is very much
possible.

\subsection{RSA: Bypass via Missing Parameters}\label{sec:mbed:RSA:missing_params}

Recalling \autoref{sec:background}, an RSA private key is composed by some
redundant parameters while at the same time not all of them are mandatory per
\rfc{8017}: \textit{``An RSA private key should be represented''}. This implies
that cryptography implementations must deal with RSA private keys that do not
contain all parameters, requiring potentially computing them on demand. Natural
questions arise:
(i) \emph{How do software libraries handle this uncertainty?}
(ii) \emph{Does this uncertainty mask SCA threats?}
Shifting focus from \openssl, the remainder of this section analyzes the open
source \mbed library in this regard.

\Paragraph{Fuzzing RSA private key loading}
Following the Triggerflow methodology,
we developed unit tests for the \mbed library,
specifically for targeting RSA key loading code paths.
To this end, we analyzed the \mbed v2.18.1 \code{bignum} implementation and set three
POIs for Triggerflow:
(i) GCD computation, \mLibGCD;
(ii) Modular multiplicative inverse, \mLibInv;
(iii) Modular exponentiation, \mLibExp.
We arrived at these POIs from state-of-the-art SCA applied
to cryptography libraries where these operations are commonly exploited.
The first two functions are based on the binary GCD algorithm, previously
shown weak to SCA
\citep{DBLP:conf/ima/AciicmezGS07,DBLP:journals/jce/AldayaSS17,DBLP:conf/uss/GarciaB17,DBLP:conf/ccs/WeiserSB18,our_ches19_paper},
while exponentiation is a classical SCA target
\citep{DBLP:conf/crypto/KocherJJ99,DBLP:conf/icics/ClavierFGRV10,DBLP:conf/ches/GenkinPPT15,DBLP:conf/ccs/GarciaBY16,DBLP:conf/ches/BernsteinBGBHLV17}.

With these POIs, we fuzz the RSA \mbed private key loading code path to
identify possible vulnerabilities. The fuzzing consists of testing
the loading of an RSA private key when some parameters are equal to zero
(i.e.\ empty \pkcsOne parameter).

After configuring the potential leaking functions as Triggerflow POIs,
we created an RSA private key fuzzing utility that generates all possible combinations
of \pkcsOne-compliant private keys.
This ranges from a private key that includes all \pkcsOne parameters to none.
While the latter is clearly invalid as it carries no information,
other missing combinations could be interesting regarding SCA.
As \pkcsOne defines eight parameters, the number of private key combinations
compliant with this standard is 256.

Triggerflow provides a powerful framework for testing all these combinations smoothly.
Using Triggerflow for each of these private keys, we tested the
generic function of \mbed for loading public keys:
\code{mbedtls\_pk\_parse\_keyfile}.
The advantage of using Triggerflow for this task is that we can automate
the whole process of testing each code corner of this execution path,
searching for SCA threats.
\autoref{fig:triggers} (bottom) shows a Triggerflow unit test
of one of these parameter combinations,
with a private key missing $d$.
Unit tests for the other combinations are similar.

\Paragraph{Results}
For each combination, we obtained a
report that indicates if and where POIs were hit or not,
also recording the program return code.
A quick analysis of the generated reports indicates
the 256 combinations group in four classes
(i.e.\ only four unique reports were generated for all 256
private key parameter combinations).
\autoref{tab:pem_fuzz_results} shows the number of keys for each group.
The majority of private key combinations yield an ``Invalid''
return code without hitting a POI before returning.

The group ``Public'' contains those remaining \emph{valid} private keys
for which $\{d, p, q\}$ is not a subset of included parameters.
In this case, \mbed recognized the key as a public key
even if the CRT secret parameters are present.
Nevertheless, identified as ``Public'' by \mbed, we ignore them,
since no secret data processing takes place.

\begin{table}[h]
	\caption{Report groups for the 256 private keys.}
	\centering
	\begin{tabular}{|c|c|}
		\hline
		Group & Number of keys  \\
		\hline
		\hline
		Invalid & 216 \\
		\hline
		Public & 8 \\
		\hline
		POI-hit (CRT) & 16 \\
		\hline
		POI-hit (CRT \& $d$) & 16 \\
		\hline
	\end{tabular}
	\label{tab:pem_fuzz_results}
\end{table}

The last two groups in \autoref{tab:pem_fuzz_results} contain those
private keys (32 in total) that indeed hit at least one POI.
Analyzing both reports on these groups, we identified two potential leakage points.
One is related to processing of the CRT parameters,
and the other to computation of the private exponent $d$.
We now investigate if these hits represent an SCA threat.
\autoref{sec:appx_mbedtls_keys} details the complete list of parameter
combinations that hit a POI.

\Paragraph{Leakage analysis: CRT}
The last two report groups have at least one hit at a Triggerflow POI
in a CRT related computation. In both groups, the report regarding
this code path is identical, hence the following analysis applies to both.

The Triggerflow report reveals hitting the modular inverse POI;
the parent function is \code{mbedtls\_\-rsa\_\-deduce\_\-crt},
computing the CRT parameters in \eqref{eq:crt_parameters} as
$i_q=q^{-1} \bmod p$ using \mLibInv.
It is a variant of the binary extended Euclidean
algorithm (BEEA) with an execution flow highly dependent on its inputs,
therefore an SCA vulnerability.
This is similar to \openssl{}'s \autoref{sec:rsa_openssl} vulnerability.
Yet in contrast to \openssl, this code path in \mbed
executes every time this library loads a private key:
the vulnerability exists regardless of missing parameters in the private key.

\Paragraph{Leakage analysis: private exponent}
The last group in \autoref{tab:pem_fuzz_results} contains the CRT
leakage previously described in addition to one related to
private exponent $d$ processing.
The targeted POIs hit by all private key parameter combinations
in this group are \mLibGCD and \mLibInv. Both are called by the
parent function \code{mbedtls\_rsa\_deduce\_private\_exponent}, that aims
at computing the private exponent if it is missing in the private key using
\eqref{eq:d}, involving a modular inversion.
However, for computing $\lcm(p-1,q-1)$ using \eqref{eq:lcm},
the value $\gcd(p-1,q-1)$ needs to be computed first.
Therefore, the report indicates a call first to \mLibGCD with inputs $p-1$ and $q-1$.
This call represents an SCA vulnerability as the binary GCD
algorithm is vulnerable in these instances \citep{DBLP:conf/ima/AciicmezGS07,DBLP:journals/ijcta/AldayaMSS17,our_ches19_paper}.
Note, this leakage is also present in \openssl (\autoref{sec:rsa_openssl}),
however the contexts differ. We observed \openssl leakage when verifying $d$
correctness, whereas \mbed computes $d$ because it is missing. This difference
is crucial regarding SCA, because \openssl verifies by checking if $de=1 \bmod
\lcm(p-1,q-1)$ holds; yet \mbed indeed computes $d$, executing a modular
inversion \eqref{eq:d}. Therefore this vulnerability is present in \mbed, and
absent in \openssl.

After obtaining $\lcm(p-1,q-1)$, it computes $d$ using \eqref{eq:d} through a
call to \mLibInv. \citep{DBLP:conf/uss/GarciaB17,DBLP:conf/ccs/WeiserSB18} exploit
\openssl{}'s BEEA using microarchitecture attacks, so at a high level it represents
a serious security threat. A deeper analysis follows for this \mbed case.

Summarizing, the private exponent computation in \mbed contains two vulnerable code paths:
(i) GCD computation of $p-1$ and $q-1$; and
(ii) modular inverse computation of $e$ modulo $\lcm(p-1,q-1)$.
Next, we investigate which of these represents the most critical threat.

The inputs of the first code path (GCD computation) are roughly the same size.
This characteristic implies that, for some SCA signals, the number of bits
that can be recovered is small and not sufficient to break RSA.
\citep{DBLP:journals/jce/AldayaSS17,DBLP:conf/uss/GarciaB17} practically demonstrated this limitation
using different SCA techniques: the former power consumption, the latter microarchitecture timings.

However, note the inputs of the second code path (modular inversion) differ
considerably in size. The public exponent $e$ is typically small, e.g.\ 65537.
Following \eqref{eq:lcm}, $\lcm(p-1,q-1)$ has roughly the same number of bits as
$(p-1)(q-1)$; more than 1024 because $\gcd(p-1,q-1)$ is small with high
probability \citep{MR2554564}.
This significant bit length difference between \mLibInv inputs makes this algorithm extremely
vulnerable to SCA \citep{DBLP:journals/ijcta/AldayaMSS17}.
This difference implies the attacker knows part of the algorithm execution flow
beforehand, and it is exactly this part that is usually difficult to obtain and
considerably limits the number of bits that can be recovered employing some SCA
techniques as demonstrated in
\citep{DBLP:journals/jce/AldayaSS17,DBLP:conf/uss/GarciaB17}.
This characteristic means the attacker only needs to distinguish the
main two arithmetic operations present in this algorithm (i.e.\ right-shift and subtraction)
to fully recover the input $\lcm(p-1,q-1)$ that yields $d$.

\begin{figure}
	\includegraphics[width=\linewidth]{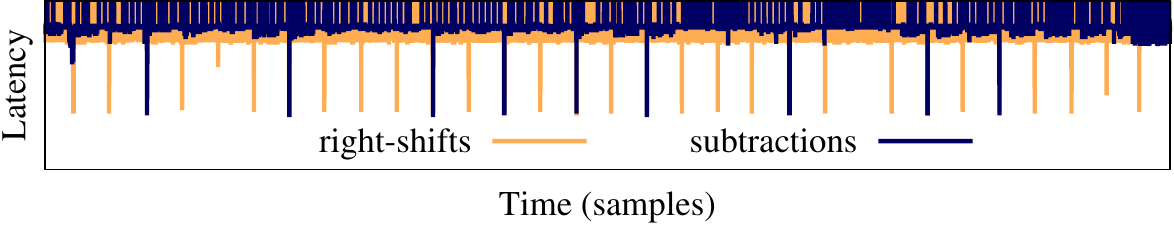}
	\caption{Sequence of right-shifts and subtractions from a \fl attack targeting \mbed modular inversion.}
	\label{fig:flush_reload_mbedtls}
\end{figure}

Regarding microarchitecture attacks, this distinction lends itself to
a \fl attack. As part of our validation, we attacked this implementation
using a \fl attack paired with a performance degradation technique \citep{DBLP:conf/acsac/AllanBFPY16}.
We probed two cache lines: one detecting right-shift executions, the other subtractions.
\autoref{fig:flush_reload_mbedtls} shows the start of a trace, demonstrating the
sequence extraction of right-shifts and subtraction is straightforward.

In addition, the key loading application threat model allows capturing
several traces corresponding to the processing of the same secret data.
Therefore, the attacker can correct errors that may appear in captured traces
(e.g.\ fix errors produced by preemptions) by combining the information as they
are redundant.

\Paragraph{Recap}
After the analysis of both leaking code paths we detected,
we conclude the private exponent leakage
is easier to exploit than that of CRT due to the large \bitlen difference
between the modular inversion algorithm inputs in the former
\citep{DBLP:journals/ijcta/AldayaMSS17,DBLP:conf/ccs/WeiserSB18,our_ches19_paper}.
On the other hand, the private exponent leakage is only present
when the private key does not include $d$;
whereas the CRT-related leakage always represents a threat regardless of missing parameters
\citep{cryptoeprint:2020:055}.
The number of bits that can be recovered exploiting these leaking code paths
depends on the side-channel signal employed.
However, these code paths potentially leak all the bits of the processed secrets,
as demonstrated in
\citep{DBLP:journals/ijcta/AldayaMSS17,DBLP:conf/ccs/WeiserSB18,our_ches19_paper,cryptoeprint:2020:055}.

\Paragraph{Keys in the wild}
As such, in the context of \mbed the simplest example of a vulnerable RSA key is
the default key typically generated by libraries, including \emph{all}
parameters. We verified this default behavior on e.g.\ \mbed, \openssl, and
BoringSSL. Hence such keys are ubiquitous in nature. For example, Let's
Encrypt's \code{certbot} tool for automated certificate renewal only supports
RSA keys. We conclude that any application linking to \mbed for RSA
functionality including key parsing is potentially vulnerable, including (but
certainly not limited to) ACME-backed web servers relying on \mbed for TLS
functionality.

\section{Two End-to-End Attacks} \label{sec:attacks}

As highlighted in \autoref{sec:vuln}, the format used to
encode a private key can lead to the bypass of side-channel
countermeasures in cryptographic libraries: these are \emph{Certified Side Channels}.
In this section we concretely instantiate the threat in \autoref{sec:expl_params_vuln} with
two SCA attacks against ECDSA signature generation over the popular
\nistp{} curve against \opensslver{}: a remote timing attack and an EM attack.

\Paragraph{Target application}
For computing the ECDSA signatures from the protocol stack application layer we chose
\rfc{3161} Time Stamp Protocol. The protocol ensures the means of establishing
a time stamping service: a time stamp request message from a client and the 
corresponding time stamp response from the Trusted Timestamp Authority (TSA).
In short, the TSA acts as a trusted third party that binds the Time Stamp Token (TST)
to a valid client request message---one way hash of some information---and digitally
signs it with the private key. Anyone with a valid TSA certificate can thus verify
the existence of the information with the particular time stamp, ensuring timeliness and non-repudiation.

In principle, the client generates a time stamp request message containing the
version information, OID of the one way hash algorithm, and a valid hash of the data.
Optionally, the client may also send TSA policy OID to be used for creating the time
stamp instead of TSA default policy, a random nonce for verifying the response time
of the server, and additionally request the signing public key certificate in the TSA
response message. The server timestamp response contains a status value and a TST
with the OID for the content type and the content itself composed of DER-encoded
TST information (TSTinfo). The TSTinfo field incorporates the version number info,
the TSA policy used to generate the time stamp response, the message imprint (same
as the hashed data in the client request), a unique serial number for the TST,
and the UTC based TST generation time along with the accuracy in terms of the time granularity.
Depending on the client request, the server response may additionally contain
the signing certificate and the client provided nonce value. For further details on
TSP, the reader may refer to \rfc{3161}.

Our attack exploits point multiplication in the ECDSA signature generation
during the TSA response phase to recover the long term private key of the
server. As a protocol-level target, we compiled and deployed unmodified
uts-server\footurl{https://github.com/kakwa/uts-server} v0.2.0 without debug
symbols, an open source TSA server linking against an unmodified debug build of
\opensslver{}. We configured the server with a \nistp{} X.509 digital
certificate, using the private key containing explicit parameters with a zero
cofactor, i.e.\ the preconditions for our \autoref{sec:expl_params_vuln}
vulnerability. We used the \openssl time stamp utility \code{ts} to create time
stamp requests with SHA256 as the hash function, along with a request for the
server's public key certificate for verification. We used the provided HTTP
configuration for uts-server, hence the TSP messages between the (victim) server
and our (attacker) client were transported via standard HTTP.

\Paragraph{Target device}
We selected a Linux-based PINE A64-LTS board with an Allwinner A64 Quad Core SoC
based on Cortex-A53 which supports a 64-bit instruction set with a maximum clock
frequency of 1.15 GHz. The board runs Ubuntu 16.04.1 LTS without any
modifications to the stock image. We set the board's frequency governor to
``performance''.

\Paragraph{Threat model}
As discussed (\autoref{sec:expl_params_vuln}), when handling such a key in
\opensslver{}, the underlying implementation for the EC scalar multiplication is
based on a \wnaf{} algorithm, which has been repeatedly targeted in SCA works
over the last decade, usually focusing on the recovery of the LSBs of the secret
scalar. Contributions from Google \citep{DBLP:conf/fc/Kasper11} partially
mitigated the attack vector for select named curves with new \code{EC\_METHOD}
implementations, then fully even for generic curves due to the results and
contributions from \citep{DBLP:conf/acsac/TuveriHGB18}.
With the attack vector now open again, this section presents two end-to-end
attacks with different signal procurement methods:
(i) a novel remote timing attack (\autoref{sec:attacks:timing}), where it is
assumed the attacker can measure the overall wall clock time it takes for the
TSA server to respond to a request---note this attacker is indistinguishable
from a legitimate user of the service;
(ii) an EM attack (\autoref{sec:attacks:EM}), similar in spirit to
\citep{DBLP:conf/ccs/GenkinPPTY16,DBLP:conf/acsac/TuveriHGB18}, which has the
same aforementioned threat model but additionally assumes physical proximity to
non-invasively measure EM emanations.
The motivation for the two different threat models is due to both practicality
and the number of required samples, which will become evident by the end of this
section.

\subsection{ECDSA: Remote Timing Attack}\label{sec:attacks:timing}

In contrast to previous work on this code path and to widen potential real-world
application, we performed a remote timing attack on the TSA server application
via TCP. Instead of taking measurements on this code path server side like e.g.\
\pp \citep{DBLP:conf/asiacrypt/BrumleyH09,DBLP:conf/cosade/Brumley15} and
\fl \citep{DBLP:conf/ches/BengerPSY14,DBLP:conf/ctrsa/PolSY15,DBLP:conf/acsac/AllanBFPY16,DBLP:conf/ccs/FanWC16},
we (as a non-priviliged, normal user of the service), make network requests and
measure the wall clock response time.

\Paragraph{Experiment setup}
We connected the PINE A64-LTS board directly by Ethernet cable to a workstation
equipped with an Intel i5-4570 CPU and an onboard I217-LM (rev 04) Ethernet
controller. To measure the remote wall clock latency and reduce noise, we
created a custom HTTP client for time stamp requests. Its algorithm is as
follows:
(i) establish a TCP connection to the server;
(ii) write the HTTP request and the body, sans a single byte;
(iii) start the timer;
(iv) write the last body byte---now the server can begin computing the digital signature;
(v) read the HTTP response headers---the server might write at least part of them
before computing the digital signature;
(vi) read one byte of HTTP response body---the digital signature is received by
the server directly from linked OpenSSL in an octet string, so reading one byte
guarantees it has been generated;
(vii) stop the timer;
(viii) finish reading the HTTP response;
(ix) record the timing information and digital signature in a database;
(x) close the TCP connection;
(xi) repeat until the requested number of samples has been gathered.
We implemented the measurement software in C to achieve maximum performance and control over 
operations.
For the client timer, we used the x86 \code{rtdtsc} instruction that is freely
accessible from user space. In recent Intel processors the \code{constant\_tsc}
feature is available---a frequency-independent and easily accessible precision
timer.

\begin{figure}
\includegraphics[width=\linewidth]{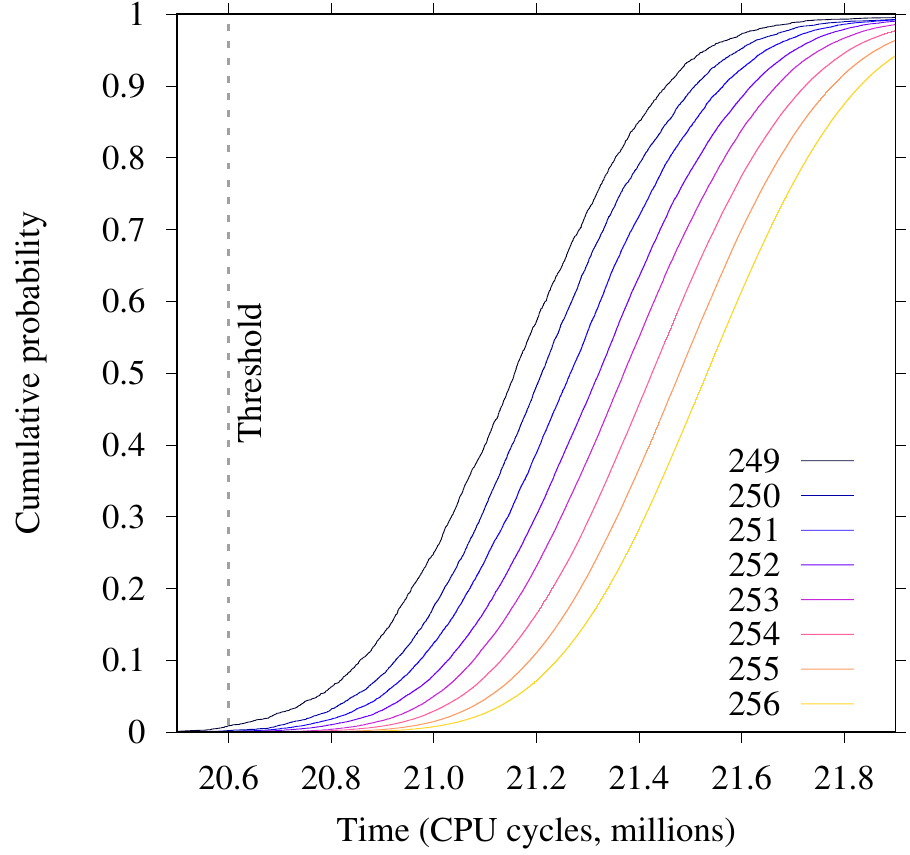}
\caption{%
    Direct correlation between wall-clock execution time of ECDSA
    signature generation and the \bitlen of the nonce.
    Plots from left to right correspond to legend keys from top to bottom.
    Measured on \nistp in \openssl on a Pine64-LTS,
    bypassing all SCA hardening countermeasures via a private key
    parsing trigger.
}%
\label{fig:timings}
\end{figure}

Performing a traditional timing analysis under the above assumptions, we
discovered a direct correlation between the wall clock
execution time of ECDSA signature generation and the \bitlen of the
nonce used to compute the signature, as shown in \autoref{fig:timings}.
This happens because given a scalar $k$ and its recoded \naf{} representation
$\hat{k}$, the algorithm execution time is a function of both the \naf{} length
of $\hat{k}$ and its Hamming weight. While the \naf{} length is a good
approximation for the \bitlen of $k$ (in fact at most one digit longer), its
Hamming weight masks the \naf{} length linearly so it is not obvious how to
correlate these two factors with the precise \bitlen of $k$. Nevertheless, the
empirical results (by sampling) shown in \autoref{fig:timings} clearly
demonstrate the latter is directly proportional to the overall algorithm
execution time.

This result shares similarity to the one exploited in
\CVE{2011-1945} \citep{DBLP:conf/esorics/BrumleyT11} (that built the foundation
for the recent Minerva\footnote{\url{https://minerva.crocs.fi.muni.cz/}}
and TPM-FAIL \cite{temp:tmpfail} attacks), and in fact suggests that CVE applied
to not only binary curves using the Montgomery ladder, but prime curves as well.
Following their attack methodology, we devise an attack in two phases:
(i) The collection phase exploits
the timing dependency between the execution time and the \bitlen of the
nonce used to generate a signature, thus selecting $(r,s,m)$
tuples associated with shorter-than-average nonces;
(ii) The recovery phase then combines the partial knowledge inferred from
the collection phase to instantiate an HNP instance and solve it through
a lattice technique (\autoref{sec:bg:lattice}).

\Paragraph{Collection phase}
Using our custom TCP time stamp client, across Ethernet
we collect $500$K traces for a single attack, sorting by the measured
latency, and filter the first $t=128$ items: empirically this is closely
related to the selection by a fixed threshold suggested by
\autoref{fig:timings}.
We prefer the formulation where we set the dimension $t$ of the filtered
set and the total number of collected signatures, as these numbers are
more significant for comparison with other works or directly used in the
formalization of the subsequent lattice phase.

\Paragraph{Lattice phase}
As noted above, the collection phase in this attack selects
shorter-than-average nonces, i.e.\ looking at the nonce $k_i$ as a string
of bits with the same \bitlen of the generator order $n$,
$$
0 < k_i < 2^{(\lg{(n)} - \ell_i)} < 2^{(\lg{(n)} - \ell)} < n/2^{\ell} \equiv n/W < n
$$
for some $W = 2^{\ell}$ bound, representing that at least $\ell$
consecutive MSBs are equal to $0$.
This is in contrast with the \autoref{sec:bg:lattice} formalization,
which instead implies knowledge of nonce LSBs,
so we need to slightly revise some definitions to frame the lattice
problem using the same notation.
Therefore, we can define $W_i = W = 2^{\ell}$ and, similarly to the
formalization in \autoref{sec:bg:lattice}, rearrange \eqref{eq:ecdsa} as
$k_i = \alpha (r_i/s_i) - (-h_i/s_i) \bmod n$ and then redefine
$t_i = \lfloor r_i/s_i \rfloor_n$, $\hat{u}_i = \lfloor -h_i/s_i \rfloor_n$
which leads once again to
$0 \leq \lfloor \alpha t_i - \hat{u}_i \rfloor_n < n/W_i$,
from which the rest of the previous formalization follows unchanged.

Although it used a different lattice description,
\citep{DBLP:conf/esorics/BrumleyT11} also dealt with a leak based on
nonce MSBs, which led to an interesting property that is valid also for
the formalization used in this particular lattice attack.
Comparing the definitions of $t_i$, $\hat{u}_i$, and $u_i$
above with the ones from \autoref{sec:bg:lattice}, we note that
in this particular attack no analogue of the $a_i$ term features in
the equations composing the lattice problem, from which follows that
even if some $k_i$ does not strictly satisfy the bound $k_i < n/W$ there
is still a chance that the attack will succeed, leading to a better resilience
to \emph{errors} (i.e., entries in the lattice that do not strictly satisfy the
bound above) in this lattice formulation. From the attacker perspective, higher
$W$ is desirable but requires more leakage from the victim.

Since in this formulation the attacker does not use a per-equation $W_i$ as the
distributions are partially overlapping, the question remains how to set $W$.
Underestimating $W$ is technically accurate for approximating zero-MSBs for most
of the filtered traces, but forces higher lattice dimensions and slower
computation for each job.
Using a larger set of training samples, analyzing the ground truth
w.r.t.\ the actual nonce of each sample, we empirically determined that
the distribution of nonce \bitlen{}s on the average set filtered by our
collection phase is a Gaussian distribution with mean $\overline{\lg{(k_i)}} = 247.80$
and s.d.\ $3.81$, which suggests $W=2^8$ ($8=256-248$) is a better approximation
of the bound on most nonces. Given the s.d.\ magnitude, by trial-and-error we
set $W=2^7$ as a good trade-off for lattice attack execution time vs.\ success
rate.

Combining the better resilience to errors of this particular lattice formulation
and the higher amount of information carried by each trace included in the
lattice instance by pushing $W$, we fixed the lattice attack parameters to
$d=60$ and $j=55$K and limit the maximum number of attempted lattice reductions
per job to 100 (in practice on our cluster, less than a single minute), as we
observed the overwhelming majority of instances returned success within this
time frame or not at all.

\Paragraph{Attack results}
With these parameters, and repeating the attack 100 times, we observed a 91\%
success rate in our remote timing attack over Ethernet. The median number of
jobs needed over all attack instances was $1377$ (i.e.\ $j=1377$ was sufficient
for key recovery in the majority of cases).
Those reductions that led to successful key recovery (i.e.\ $91$ in number) had
$\overline{\lg{(k_i)}} = 246.85$ and s.d.\ $3.13$, while the $j=55$K reductions
per \emph{each} of the 9 failed overall attack instances had $\overline{\lg{(k_i)}} = 247.96$
and s.d.\ $3.87$.
This difference suggests:
(i) the better resilience to errors in this lattice formulation is empirically
valid, as given the stated s.d.\ not all $k_i$ satisfied the bound $W$;
(ii) in our environment, even the failed instances would likely succeed by
tweaking lattice parameters (i.e.\ decreasing $W$ and increasing $d$) and
providing more parallel computation power (i.e.\ increasing $j$).

In case of success, the attacker obtains the long-term secret key. On
failure she can repeat the collection phase (accumulating more traces
and improving the filtering output and the probability of success of
another lattice phase) or iteratively tune the lattice parameters
(decreasing $W$ and increasing $d$) to adapt to the features of the
specific output of the collection phase, thus improving the lattice attack's
success probability.
\subsection{ECDSA: EM Attack}\label{sec:attacks:EM}

In a much stronger (yet still SCA-classical) attack model assuming physical
proximity, we now perform an EM attack on \openssl ECDSA.
As far as we are aware, we are the first to exploit this code path in the context of OpenSSL and \nistp{}:
\citep{DBLP:conf/ccs/GenkinPPTY16} target the 256-bit Bitcoin curve, and
\citep{DBLP:conf/acsac/TuveriHGB18} the 256-bit SM2 curve.
The reason for this is our \autoref{sec:expl_params_vuln} vulnerability allows
us to bypass the dedicated \code{EC\_METHOD} instance on this architecture,
\code{EC\_GFp\_nistz256\_method} which is constant time and optimized for AVX
and ARMv8 architectures.
The wNAF Double and Add operations have a different set of underlying
finite field operations---square, multiply, add, sub, inversion---resulting in distinguishable
EM signatures.

\begin{figure}
\includegraphics[width=\linewidth]{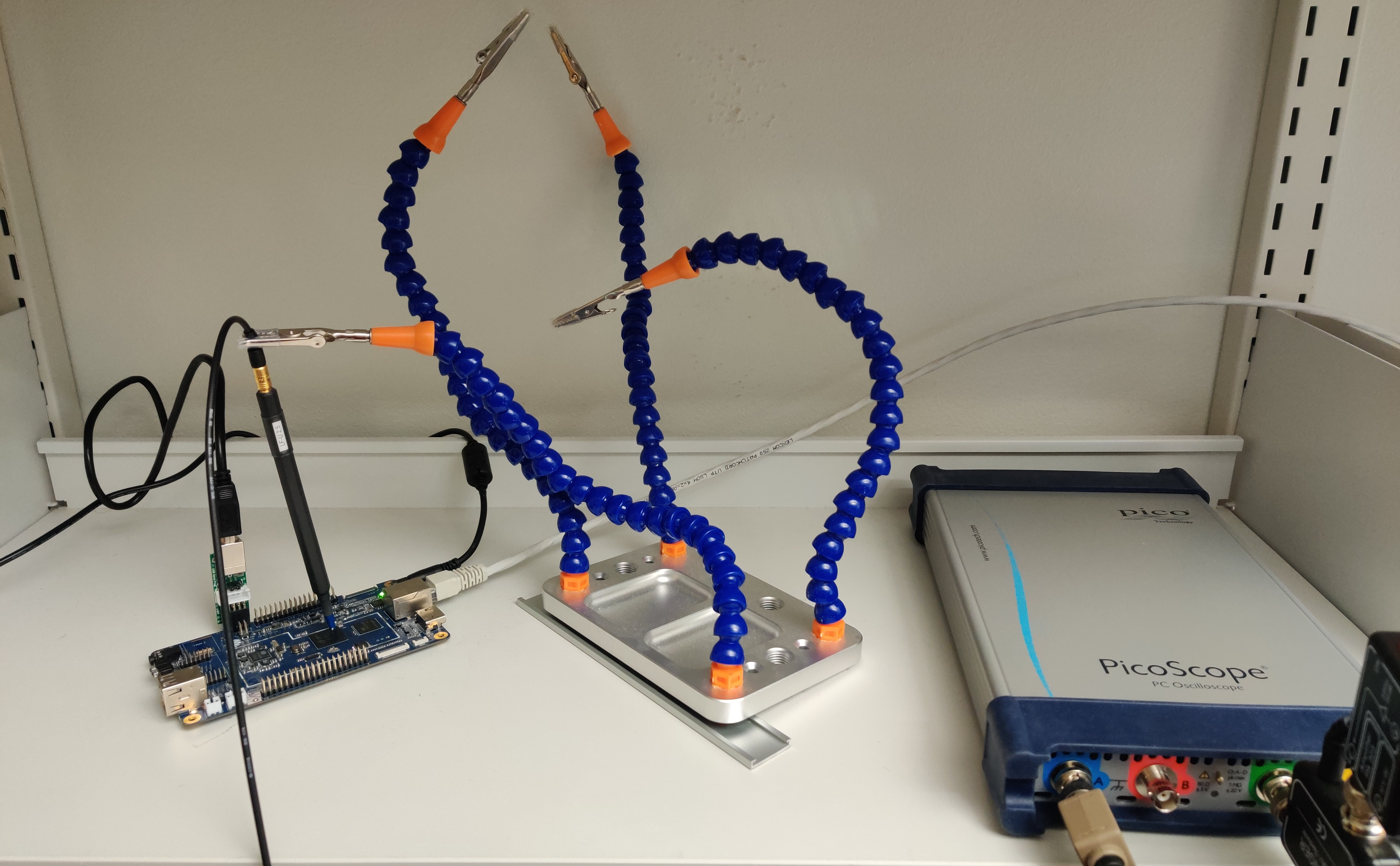}
\caption{Experiment setup capturing EM traces using Picoscope USB oscilloscope
with the Langer EM probe positioned on the Pine64-LTS SoC: a TSP server
connected via Ethernet serving requests over HTTP.}
\label{fig:EM_setup}
\end{figure}

\Paragraph{Experiment setup}
To capture the EM traces, we positioned the Langer LF-U $2.5$ near field probe head
on the SoC where it resulted in the highest signal quality. For digitizing the EM
emanations, we used Picoscope $6404C$ USB digital oscilloscope with a bandwidth of
500 MHz and maximum sampling rate of 5 GSps. However, we used a lower sampling rate of
125 MSps as the best compromise between the trace quality and processing
overhead. To acquire the traces while ensuring that the entire ECDSA trace was
captured, we synchronized the oscilloscope capture with the time stamp request
message: initiate the oscilloscope to start acquiring traces,
query a time stamp request over HTTP to the server and wait for the server response,
and finally stop the trace acquisition. We stored the EM traces along with the DER-encoded
server response messages. We parsed the messages to retrieve the hash from the
client request and the DER-encoded ECDSA signatures, used to generate metadata for the
key recovery phase. \autoref{fig:EM_setup} shows the setup we used for our attack.

\Paragraph{Signal analysis}
After capturing the traces, we moved to offline post processing of the EM traces
for recovering the partial nonce information. This essentially means identifying the position
of the last Add operation. The problem is twofold: finding the end of the point
multiplication (end trigger), then identifying the last Add operation therein.
We divided the complete signal processing phase mainly into four steps:
(i) Remove traces with errors due to acquisition process;
(ii) Find the end of the ECDSA point multiplication;
(iii) Remove traces encountering interrupts;
(iv) Identifying the position of the last Add operation.
We started by selecting only those traces which had peak magnitude to the root mean square ratio
within an emphatically selected confidence interval, evidently removing traces where the point
multiplication operation was not captured or trace was too noisy to start with.

In the next step, we used a specific pattern at the end of ECDSA point multiplication
as our soft end trigger. To isolate
this trigger pattern from the rest of the signal, we first applied a low pass FIR filter followed
by a phase demodulation using the digital Hilbert transform. We further enhanced this
pattern while suppressing the rest of the operations by applying root mean square
envelope with a window size roughly half its sample size. We created a template
by extracting this pattern from 20 random traces and taking their average. We used the Euclidean distance
between the trace and template to find the end of point multiplication. We dropped all traces
where the Euclidean distance was above an experimental threshold value, i.e.\ no soft
trigger found.
The traces also encountered random interrupts due to OS scheduling clearly
identifiable as high amplitude peaks. Any traces with an interrupt at the end of
point multiplication were also discarded to avoid corrupting the detection of the Add operation.

\begin{figure}
\includegraphics[width=\linewidth]{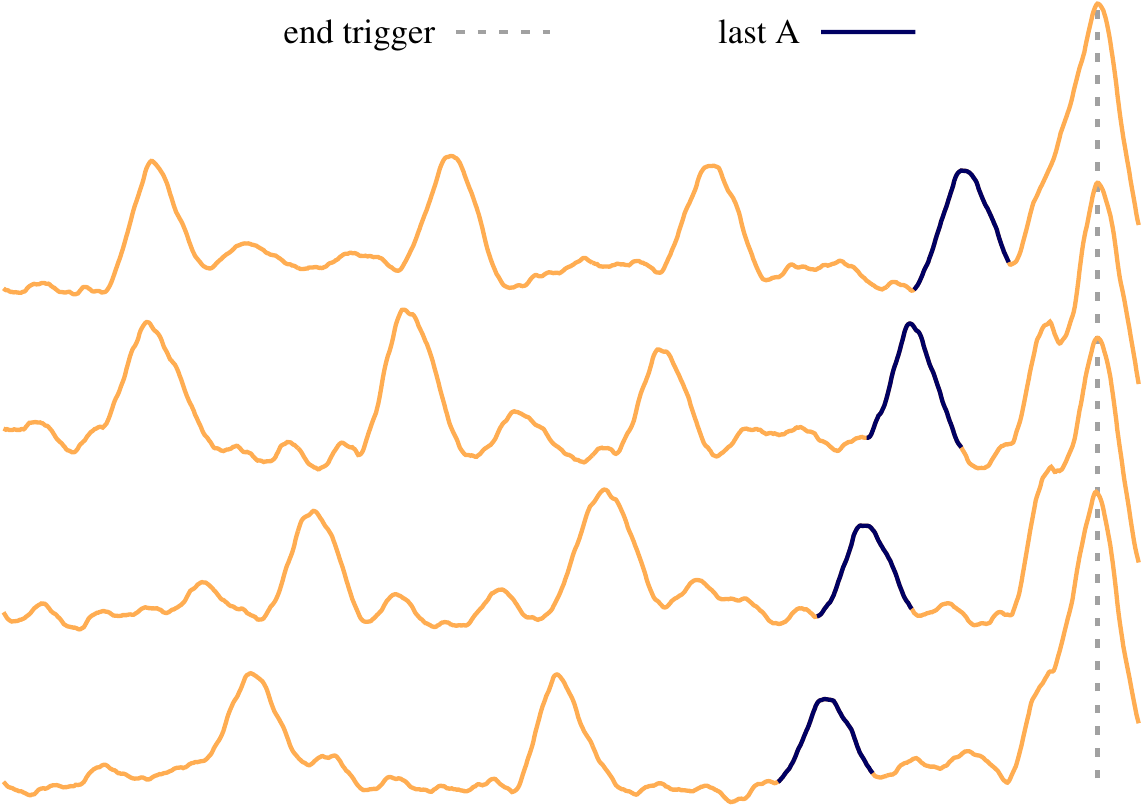}
\caption{Four different EM traces showing the last Add (A) operations relative to the soft end trigger.
         The distance in terms of samples between the last Add and trigger gives
         the number of Double (D) operations. Top to Bottom: Trace ends with an A, AD, ADD, ADDD.}
\label{fig:trace_em}
\end{figure}

To recover the position of the last Add operation, we applied a different set of filters
on the raw trace, keeping the end of point multiplication as our starting reference.
Since the frequency analysis revealed most of the Add operations energy is between 40 MHz and 50 MHz,
we applied a band pass FIR filter around this band. Performing a digital
Hilbert transform, additional signal smoothing and peak envelope detection, the Add
operations were clearly identifiable (\autoref{fig:trace_em}).

To automatically extract the Add operation, we first used peak extraction.
However it was not as reliable since the signals occasionally encountered noisy peaks or
in some instances the Add peaks were distorted. We again resorted to the template matching
method used in the previous step, i.e.\ create an Add template and use Euclidean distance
for pattern matching. For each peak identified, we also applied the template matching
and measured the resulting Euclidean distance against a threshold value. Anything greater than
the threshold was considered a false positive peak.

These steps ensured that the error rate stays low,
consequently increasing the success rate of the key recovery lattice attack.
We estimated the number of Double operations using the total sample length
from the middle of the last Add operation to the end of trace as illustrated in \autoref{fig:trace_em}.
To effectively reduce the overlap between the sample length metric of different Double
and Add sequences, we applied K-means clustering to keep sequences which were
close to the cluster mean.

\Paragraph{Attack results}
We acquired a total of 500 signatures, and after performing the signal
processing steps we were left with 422 traces.
Additionally, after filtering out signatures categorized as ``A'' and hence not
useful lattice-wise, we were left with $t=172$ signatures suitable for building
lattice problem instances. We chose $d=120$ as the number of signatures to
populate the lattice basis. We then constructed $j=48$ instances of the lattice
attack, randomly selecting $d$-size subsets from the $t$ signatures for each
instance. We then ran these instances in parallel on a 2.10 GHz dual CPU Intel
Xeon Silver 4116 (24 cores, 48 threads across 2 CPUs). The first
instance to succeed in recovering the private key did so in just over three
minutes. Checking the ground truth afterwards, $e=4$ out of the $t$ signatures
were categorized incorrectly, for a suitably small error rate of about 2.3\%.
\section{Conclusion} \label{sec:conclusion}

In this work, we evaluated how different choices of private key formats
and various optional parameters supported by them can influence
SCA security. We employed the automated tool Triggerflow to analyze
vulnerable code paths in well known cryptographic libraries for
various combinations of key formats and optional parameters.
The results uncovered several \emph{Certified Side Channels}, circumventing SCA hardened code paths in
\openssl (ECC with explicit parameters, DSA with MSBLOB and
PVK formats, RSA during key validation) and \mbed (RSA with
missing parameters). To demonstrate the severity of these vulnerabilities,
we performed microarchitecture leakage analysis on RSA and DSA and also
presented end-to-end key recovery attacks on \openssl ECDSA using traditional
timing and EM side channels.
We publish our data set for the remote timing attack to support
Open Science \citep{zenodo:2020:certified}.

In the \openssl case, \citet{DBLP:conf/ccs/GarciaBY16} conclude the fundamental
design issue around \flag is due to its insecure default nature, hypothesizing
inverted logic with secure-by-default behavior provides superior assurances.
While that would indeed have prevented \CVE{2016-2178}, our work shows that
runtime secure-by-default is still not enough: simply the presence of known
SCA-vulnerable code alongside SCA-hardened code poses a security threat. For
example, in this light, in our \autoref{sec:expl_params_vuln} vulnerability the
zero cofactor masquerades as a virtual \flag, since the exploited code path is
oblivious to the flag's value by design.

\Paragraph{Mitigations}
As part of the responsible disclosure process, we notified \openssl and \mbed
of our findings. At the same time, we made several FOSS contributions to help
mitigate these issues in \openssl, who assigned \cveours based on our work.
For the \autoref{sec:expl_params_vuln} vulnerability, we implemented a fix that
manually computes the cofactor from the field cardinality and generator order
using the Hasse Bound. This works for all standards-compliant curves---named or
with explicit parameters.
To mitigate the vulnerabilities in \autoref{sec:dsa_openssl} and
\autoref{sec:rsa_openssl}, we submitted simple patches that set the \flag
correctly, steering the computations to existing SCA-hardened code.
Moreover, we replaced the variable-time GCD function in OpenSSL by
a constant-time implementation%
\footnote{\url{https://github.com/openssl/openssl/pull/10122}}
based on the work by \citet{DBLP:journals/tches/BernsteinY19}.
To further reduce the SCA attack surface, we implemented changes%
\footurl{https://github.com/openssl/openssl/pull/9808}
in the way \openssl creates EC key abstractions when the associated curve is
defined by explicit parameters. The explicit parameters are matched against the
internal table of known curves, switching to an internal named curve
representation for matches, ultimately enabling the use of specialized
implementations where available.
Finally, we integrated the new Triggerflow unit tests (\autoref{fig:triggers}).
Applying all these fixes across non-EOL \openssl branches as well as the
development branch, no Triggerflow POIs are subsequently triggered, indicated
the patches are effective.

Astute readers may notice the above fixes do not remove the vulnerable functions
in question. In general, indeed this is one option, but such a strategy requires
analysis on a case-by-case basis. These are real-world products that come with
real-world performance constraints. For example, an SCA-secure modular
exponentiation function that protects both the length and values of the exponent
can likely meet the performance requirements for e.g.\ DSA signing, but not RSA
verification with a short, low-weight, and public exponent. This is the main
reason why libraries often feature multiple versions of the same functionality
with different security characteristics.

\Paragraph{Future work}
In \autoref{sec:attacks:timing}, discussing the lattice formulation for our
attack, we highlight an increased resilience to lattic errors compared to
similar previous works.
We note here that an analysis of error resilience of different lattice
constructions and of strategies to increase the overall success
rate of lattice attacks in the presence of errors in collected traces would
constitute a valuable future contribution to this area of research.

Our vulnerabilities in \autoref{sec:vuln} cover only a very small subset of
possible inputs, combinations, architectures, and settings. Another interesting
research question is how to extend test coverage in a reasonable way, even
considering other libraries.

\Paragraph{Acknowledgments}
We thank Tampere Center for Scientific Computing (TCSC) for generously granting
us access to computing cluster resources.
The second author was supported in part by the Tuula and Yrj\"o Neuvo Fund
through the Industrial Research Fund at Tampere University of Technology.
The third author was supported in part by a Nokia Scholarship from the Nokia
Foundation.
This project has received funding from the European Research Council (ERC) under
the European Union's Horizon 2020 research and innovation programme (grant
agreement No 804476).

\bibliographystyle{plainnat}

\appendix
\section{mbedTLS vulnerable RSA keys}\label{sec:appx_mbedtls_keys}

\begin{table}[!h]
	\small
	\newcommand{\yes}{}
	\newcommand{\no}{\textbullet}
	\caption{RSA keys that follow leaking mbedTLS code paths.
			Missing parameters are marked with \no.
			Note, the first row belongs to a key with all included parameters
			indicating that it leaks through CRT computation.}
	\centering
	\begin{tabular}{|c|c|c|c|c|c|c|c|c|}
		\hline
		Group & $N$ & $e$ & $p$  & $q$  & $d$ & $d_p$ & $d_q$ & $i_q$ \\
		\hline
		\hline
		\multirow{16}{*}{CRT} & \yes & \yes & \yes & \yes & \yes & \yes & \yes & \yes \\
		\cline{2-9}
		 & \yes & \yes & \yes & \yes & \yes & \no  & \no  & \no  \\ %
		\cline{2-9}
		 & \yes & \yes & \yes & \yes & \yes & \no  & \no  & \yes \\ %
		\cline{2-9}
		 & \yes & \yes & \yes & \yes & \yes & \no  & \yes & \no  \\ %
		\cline{2-9}
		 & \yes & \yes & \yes & \yes & \yes & \no  & \yes & \yes \\ %
		\cline{2-9}
		 & \yes & \yes & \yes & \yes & \yes & \yes & \no  & \no  \\ %
		\cline{2-9}
		 & \yes & \yes & \yes & \yes & \yes & \yes & \no  & \yes \\ %
		\cline{2-9}
		 & \yes & \yes & \yes & \yes & \yes & \yes & \yes & \no  \\ %
		\cline{2-9}
		 & \no  & \yes & \yes & \yes & \yes & \yes & \yes & \yes \\ %
		\cline{2-9}
		 & \no  & \yes & \yes & \yes & \yes & \no  & \no  & \no  \\ %
		\cline{2-9}
		 & \no  & \yes & \yes & \yes & \yes & \no  & \no  & \yes \\ %
		\cline{2-9}
		 & \no  & \yes & \yes & \yes & \yes & \no  & \yes & \no  \\ %
		\cline{2-9}
		 & \no  & \yes & \yes & \yes & \yes & \no  & \yes & \yes \\ %
		\cline{2-9}
		 & \no  & \yes & \yes & \yes & \yes & \yes & \no  & \no  \\ %
		\cline{2-9}
		 & \no  & \yes & \yes & \yes & \yes & \yes & \no  & \yes \\ %
		\cline{2-9}
		 & \no  & \yes & \yes & \yes & \yes & \yes & \yes & \no  \\ %
		\cline{2-9}
		\hline
\multirow{16}{*}{CRT \& $d$}  & \yes & \yes & \yes & \yes & \no & \yes & \yes & \yes \\ %
\cline{2-9}
& \yes & \yes & \yes & \yes & \no & \no  & \no  & \no  \\ %
\cline{2-9}
& \yes & \yes & \yes & \yes & \no & \no  & \no  & \yes \\ %
\cline{2-9}
& \yes & \yes & \yes & \yes & \no & \no  & \yes & \no  \\ %
\cline{2-9}
& \yes & \yes & \yes & \yes & \no & \no  & \yes & \yes \\ %
\cline{2-9}
& \yes & \yes & \yes & \yes & \no & \yes & \no  & \no  \\ %
\cline{2-9}
& \yes & \yes & \yes & \yes & \no & \yes & \no  & \yes \\ %
\cline{2-9}
& \yes & \yes & \yes & \yes & \no & \yes & \yes & \no  \\ %
\cline{2-9}
& \no  & \yes & \yes & \yes & \no & \yes & \yes & \yes \\ %
\cline{2-9}
& \no  & \yes & \yes & \yes & \no & \no  & \no  & \no  \\ %
\cline{2-9}
& \no  & \yes & \yes & \yes & \no & \no  & \no  & \yes \\ %
\cline{2-9}
& \no  & \yes & \yes & \yes & \no & \no  & \yes & \no  \\ %
\cline{2-9}
& \no  & \yes & \yes & \yes & \no & \no  & \yes & \yes \\ %
\cline{2-9}
& \no  & \yes & \yes & \yes & \no & \yes & \no  & \no  \\ %
\cline{2-9}
& \no  & \yes & \yes & \yes & \no & \yes & \no  & \yes \\ %
\cline{2-9}
& \no  & \yes & \yes & \yes & \no & \yes & \yes & \no  \\ %
\hline
	\end{tabular}
	\label{tab:mbedtls_vuln_rsa_keys}
\end{table} 
\end{document}